# Passivating Surface Defects and Reducing Interface Recombination in CuInS$_2$ Solar Cells by a Facile Solution Treatment

*Mohit Sood, Alberto Lomuscio, Florian Werner, Aleksandra Nikolaeva, Phillip J. Dale, Michele Melchiorre , Jérôme Guillot, Daniel Abou-Ras, Susanne Siebentritt\**

M. Sood, Dr. A. Lomuscio, Dr. F. Werner, Prof. P.J. Dale, Dr. M. Melchiorre, Prof. S. Siebentritt

Department of Physics and Materials Science, University of Luxembourg, Belvaux, L-4422, Luxembourg

E-mail: susanne.siebentritt@uni.lu

*Dr. J. Guillot*

Luxembourg Institute of Science and Technology, Belvaux, L-4422, Luxembourg

A. Nikolaeva, Dr. D. Abou-Ras

Helmholtz-Zentrum Berlin für Materialien und Energie GmbH, Hahn-Meitner-Platz 1, 14109 Berlin, Germany

**Abstract**

Interface recombination at the absorber buffer interface impedes the efficiency of a solar cell with an otherwise excellent absorber. The internal voltage or the quasi-Fermi level splitting (qFLs) measures the quality of the absorber. Interface recombination reduces the open circuit voltage (V$_{OC}$) with respect to the qFLs. The present work explores a facile sulfur-based post-deposition treatment (S-PDT) to passivate the interface of CuInS$_2$ thin films grown under Cu-rich conditions, which show excellent qFLs values, but much lower V$_{OC}$s. The CuInS$_2$ absorbers are treated in three different S-containing solutions at 80 ºC. Absolute calibrated photoluminescence and current-voltage measurements demonstrate a reduction of the deficit between qFLs and V$_{OC}$ in the best S-PDT device by almost one third compared to the untreated device. Analysis of temperature dependence of the open-circuit voltage shows increased activation energy for the dominant recombination path, indicating less interface recombination. In addition, capacitance transient measurements reveal the presence of slow metastable defects in the untreated solar cell. The slow response is considerably reduced by the S-PDT, suggesting passivation of these slow metastable defects. The results demonstrate the effectiveness of solution based S-treatment in passivating defects, presenting a promising strategy to explore and reduce defect states near the interface of chalcogenide semiconductors.

## 1. Introduction

The copper indium gallium disulfide Cu(In,Ga)S$_2$ alloy system is a promising candidate for top cell in a thin film tandem solar cell.[1] So far, a stable efficiency of 15.5% has been achieved by growing absorber at a temperature above 550 °C.[2] CuInS$_2$, the ternary compound allows to reduce additional effects due alloy disorder and band gap gradients introduced by addition of gallium.[3-5] The CuInS$_2$ absorbers grown under Cu-excess conditions exhibit higher quasi-Fermi level splitting (qFLs) compared to the absorbers grown under Cu-deficient conditions.[6] The qFLs represents the open-circuit voltage the absorber itself can produce under illumination. This qFLs is still significantly lower (~700meV) than the bandgap (1.5eV) of the absorber, particularly due to the presence of deep defects.[6,7] Moreover, solar cells realized with Cu-rich ([Cu]/[In] at % >1) absorbers suffer from large open-circuit voltage (V$_{OC}$) deficit compared to corresponding qFLs. Severe interface recombinations at the absorber/buffer (i.e. CuInS$_2$/CdS) interface are the prominent cause for this deficit.[8,9]

Interface recombination has been identified as a limiting factor in many thin film solar cells: perovskites,[10] all chalcopyrites grown under Cu-excess,[11] CdTe.[12] In that case the open circuit voltage (V$_{OC}$) of the solar cell is lower than the qFLs of the absorber. Dominating interface recombination is caused by a cliff-type band offsets *i.e.* conduction band minimum (CBM) of absorber is higher than CBM of buffer or/and by a high density of defects at or near the interface.[13] In Cu-rich $CuInS_2$ solar cells both factors can play a role: an unfavorable cliff conduction band offset between $CuInS_2$/CdS and a large number of near surface defects in the absorber.[8,14-16] The use of an appropriate buffer layer circumvents the problem of unfavorable conduction band offset at the absorber buffer interface.[17,18] However, even with a suitable band alignment, the Cu-rich sulfide devices are still dominated by interface recombination.[19] Therefore, a suitable technique to passivate the surface defects is needed.

Recent photoluminescence (PL) studies on $CuInS_2$ and $CuInSe_2$ demonstrate that the defect chemistry in both the systems is similar, establishing a close resemblance between the two systems.[1,20] Furthermore, it has been demonstrated that both selenides and sulfide chalcopyrite solar cells are dominated by interface recombination when Cu-rich absorbers are used.[9] In a recent study the dominating interface recombination was traced back to a defect present near the surface, which is related to a Se deficit.[21] This defect is caused by etching the $Cu_{2-x}Se$ secondary phase, which is always present in chalcopyrite grown under Cu-excess.[22] This defect is responsible for the $V_{OC}$ loss in Cu-rich solar cells. A similar defect is expected in the $CuInS_2$ compound. This defect affects the device $V_{OC}$ in a similar way as its counterpart $CuInSe_2$. Thus, the present work aims to find a treatment that can passivate this S deficit related defect.

In preliminary experiments, two devices were fabricated with a CdS buffer layer using low and high thiourea ($CH_4N_2S$) concentrations (i.e. the source of sulfur $S^{2-}$ ions in the chemical bath solution), where the low concentration is our standard CdS recipe. The higher thiourea concentration led to a device with higher $V_{OC}$. Since the CdS buffer layer is known to have an unfavorable band alignment with $CuInS_2$,[14,16] an additional device with Zn(O,S) buffer layer was fabricated for comparison. Details of the process for both buffer layer depositions can be found in the supplementary information. It is worth mentioning that the concentration of thiourea in the Zn(O,S) buffer layer recipe (0.4 M), is eight times more concentrated than the standard CdS buffer layer recipe (see supplementary information).

The current density-voltage (J-V) characteristics of the devices with different buffer layers and thiourea concentrations are shown in Figure S1 in the supplementary information, which show a clear improvement in device performance, especially the $V_{OC}$, with the higher thiourea concentration in the chemical bath. This improvement suggests the effect of sulfur concentration on the $V_{OC}$ of the devices. It is therefore hypothesized here that a dedicated sulfur treatment for $CuInS_2$ might be beneficial to reduce the interface recombination and improve device $V_{OC}$.

This study reports a post-deposition sulfur-treatment (S-PDT) for Cu-rich $CuInS_2$ absorbers. For PDT first the secondary phase $Cu_{2-X}S$ are etched from Cu-rich absorber using 10% KCN for 5 minutes followed by the S-PDT, i.e. the immersion of the absorbers in either ammonium sulfide (AS) or sodium sulfide (NaS) or thiourea (TU). Some of the absorbers were again etched with 5% KCN solution for 30 seconds. Finally, the absorbers are covered with buffer (Zn(O,S)) and window (aluminum doped zinc oxide i.e. AZO). Figure 1 depicts the entire procedure. These solutions were chosen because they were used for surface passivation treatments on selenide absorbers in the past,[23-25] they are used in solution processing of solar cells as a part of buffer solutions,[17,26] and because all of them contain sulfur species. The treatment aims at passivating surface defects related to the sulfur vacancy at or near the interface. We will demonstrate that S-PDT improves the device $V_{OC}$ and FF, and reduces the interface recombination, as confirmed by temperature-dependent current density-voltage (JVT) analysis.

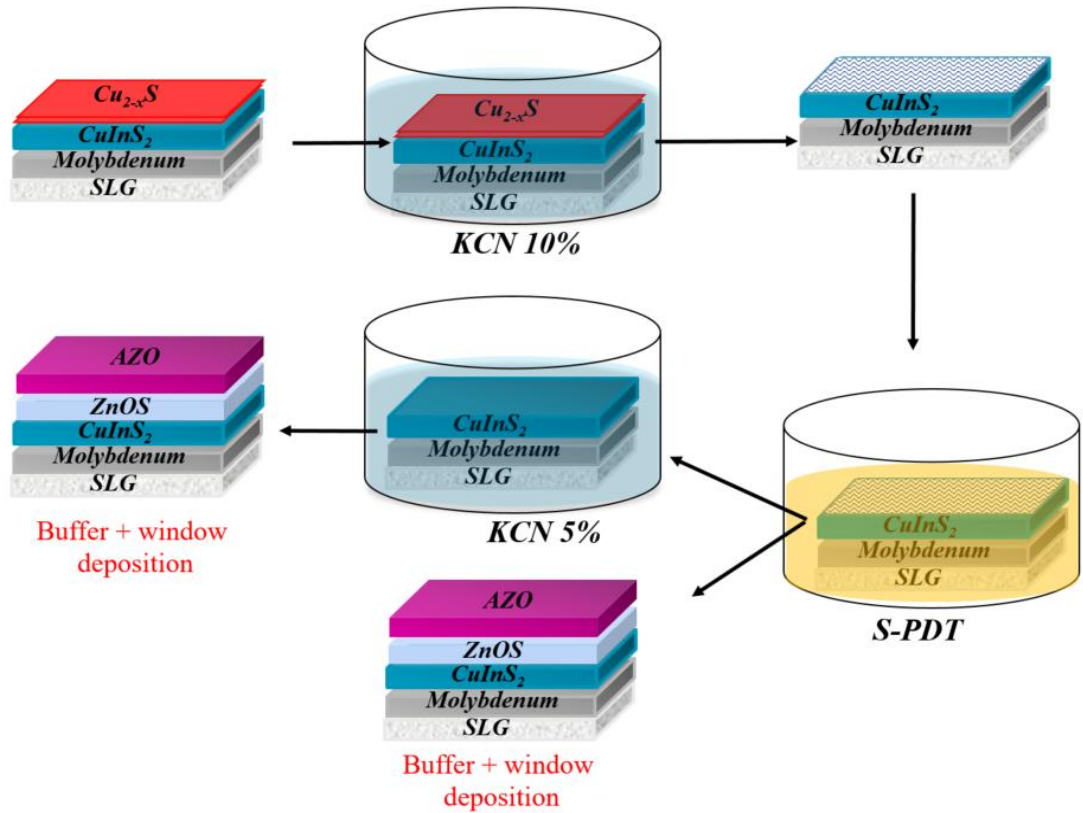

*Figure 1. A schematic diagram showing the procedure used for S-PDT.*

## 2. Changes of optoelectronic properties with S-treatments

### 2.1 Quasi-Fermi level splitting measurements

QFLs is measured by absolute calibrated PL.[27,28] Figure 2(a) depicts the transformed PL spectra transformed using Planck's generalized law and the fit to extract the qFLs, measured under 5-sun illumination.[27] A bar chart of qFLs values for the untreated and S-PDT absorbers with and without a buffer layer is presented in Figure 2(b). The first observation on untreated absorbers is that the buffer reduces qFLs, i.e. increases non-radiative recombination. This has been observed for all types of buffers that were tested in our lab [CdS, Zn(O,S), ZnMgO, not shown here]. Since, contact is necessary to make the absorber into a solar cell, it is imperative to study and improve the qFLs in absorbers covered with a buffer. Obviously, the buffer on sulfide absorbers increases recombination. This observation is in contrast to selenide absorbers, where the buffer layer was observed to passivate the surface.[28,29]

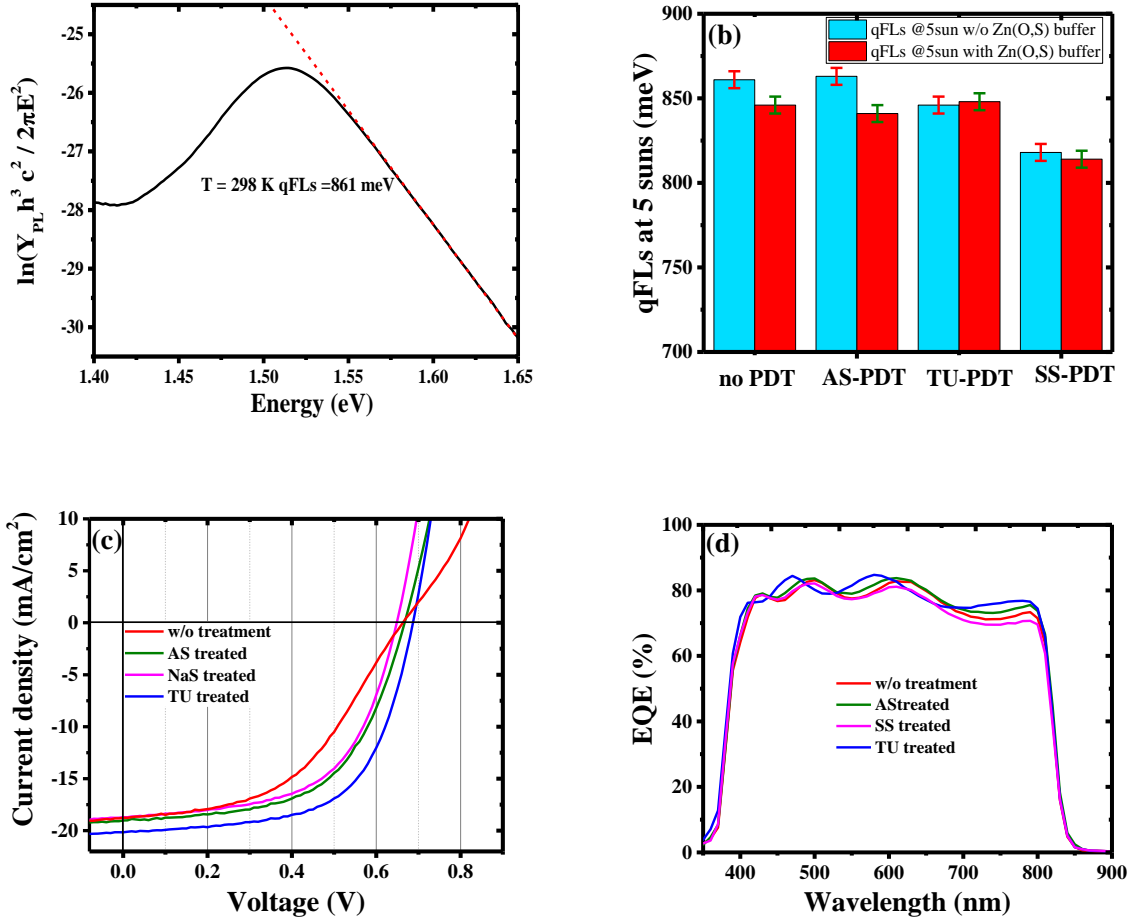

*Figure 2. (a) Exemplary PL spectra after approximation and transformation according to Planck's generalized law for the CuInS$_2$ device used to determine qFLs. (b) Quasi-Fermi level splitting values of Cu-rich CuInS$_2$ absorber treated with AS-PDT, NaS-PDT and TU-PDT, and without any PDT under 5 sun illumination with and without Zn(O,S) buffer. (c) J-V characteristics of Cu-rich CuInS$_2$ device with CuInS$_2$ absorber treated with AS, NaS and TU, and reference sample without any treatment. All samples with ZnOS buffer. (d) External quantum efficiencies of the CuInS$_2$ devices without any treatment and with AS-PDT, NaS-PDT and TU-PD.*

If we first compare the bare absorbers without buffers, no change in recombination activity is observed after AS-PDT, whereas, NaS-PDT and TU-PDT reduce the qFLs, i.e. increase non-radiative recombination in bare absorbers. The reduction in qFLs in case of NaS-PDT is significantly more than TU-PDT. This can be a result of mechanical degradation of the absorber as during the treatment partial flaking of absorber from the Molybdenum surface was observed. However, by comparing the qFLs of absorbers with and without buffer, it becomes obvious that the NaS-PDT and TU-PDT prevent the degradation due to the buffer within measurement error. And the highest qFLs with buffer is obtained for the TU treated absorber. The difference in recombination activity could be due to an improved interface or due to improved grain boundaries. To investigate if the S-treatment has an influence on the recombination activity of grain-boundaries, the best treatment (TU-PDT) was explored by cathodoluminescence. However, no difference was observed between the cathodoluminescence of the untreated and the TU-PDT absorber (Figure S3). Thus, we conclude that the main effect of the treatment is not a grain boundary passivation, but a passivation at or near the buffer/absorber interface. We investigate this further by electrical characterization of complete devices.

**2.2 Current-voltage characteristics**

Figure 2(c) shows the J-V characteristics of all treated devices together with the untreated device. The devices are as-treated, without etching after the treatment. The solar cell parameters are summarized in table 1, together with the shunt resistance determined from the slope of the J-V curve in reverse bias. We also list the qFLs values determined from the measurement at 5 suns on the samples with the ZnOS buffers and corrected for 1 sun illumination as explained in SI. For the untreated device the J-V curve exhibits an atypical 'S shape' which results in a particularly low fill factor (FF). The presence of defects near the surface are the origin of this 'S shaped' J-V curve, as will be discussed in the next section. Compared to the untreated device, none of the S-PDT devices exhibit the 'S shape' in the J-V curves. Consequently, these devices exhibit higher FF and efficiency compared to untreated devices. The S-PDT devices also exhibit slightly improved short-circuit current density ($J_{sc}$) except for NaS-PDT device, which is also the one that was mechanically damaged. To better understand the short-circuit currents, we study external quantum efficiency (EQE) spectra for the S-PDT and the untreated devices [Figure 2(d)]. All devices show a lower response in the long wavelengths region. It can be assumed that the space charge region (SCR) width is rather narrow, due to the high doping > 1e17 cm$^{-3}$ observed in Cu-rich CuInS$_2$ devices.[30] Obviously the diffusion length in these devices is not long enough to compensate for the narrow SCR. As a result there is an incomplete collection of the photons in the long-wavelength region. The lowest long wavelength response is observed in the NaS-treated and potentially mechanically damaged device. In contrast, AS-PDT and TU-PDT leads to a slight improvement in the long-wavelength region of the EQE spectra, suggesting improved diffusion length or space charge region width after the treatment. Additionally, optical effects may also play a role, as seen by the shifts in the peak wavelengths of the interference maxima. This is most evident in the EQE spectrum of the TU-PDT device, which is most distinctive among all spectra. The interference pattern in this curve is shifted to lower wavelengths [see Figure 2(d)]. This shift in interference could be due to either a thinner buffer/window (B/W) stack compared to other devices or due to a change of the optical properties of the absorber surface due to the S-PDT. Investigation of SEM cross-section on the devices (Figure S4) however shows very similar buffer-window thickness with an average value of 580 nm in all the devices. The differences between devices are smaller than the uncertainty of the thickness determination. This observation is expected, since all the devices were processed in the same window deposition run. This eliminates the possibility of a thinner B/W in TU-treated device, and suggest the modification of the optical properties of the absorber surface, either by a modification of the surface chemistry or by deposition of an additional layer.

Among all the S-PDT devices, only the TU-PDT device shows an additional improvement in the $V_{OC}$, consequently exhibiting the highest PCE of 8.5%. The $V_{OC}$ of the NaS-treated device is even lower than the one of the untreated device. In the diode model the $V_{OC}$ can be influenced by two main effects: the shunt resistance ($R_{sh}$) and the dark saturation current density ($J_o$) as follows:[31]

$$V_{OC} = \frac{Ak_BT}{q} * \ln\left(\frac{J_{sc}}{J_0} - \frac{V_{OC}}{J_0 R_{sh}}\right) \quad (1)$$

where q is the elementary charge, $k_BT/q$ is the thermal voltage, A is the diode factor and $J_{sc}$ is the photo current density which is assumed voltage-independent. The devices presented here have rather low shunt resistances (see table 1), yet, the impact of $R_{sh}$ on $V_{OC}$ is almost negligible (see discussion in SI). Thus, the differences in $V_{OC}$ are due to differences in non-radiative recombination. A comparison of $J_o$ is not possible, because the fit of the J-V characteristics to the 1-diode model is problematic, as the J-V curve does not show ideal diodic curve (more details in SI). However, the reduction in non-radiative recombination is supported by comparing the qFLs values in table 1 and Figure 2(b): the NaS-treated sample has the lowest qFLs, and the TU treated one the highest, although not significantly higher. Still, it can be concluded from the combined observation of the trends in $V_{OC}$ and qFLs: the non-radiative recombination in the TU-treated device in reduced.

*Table 1.Characteristic of best Cu-rich CuInS$_2$ device with different treatment plus reference device along with the extrapolated qFLs values at 1 sun measured on absorbers with Zn(O,S) buffer layer. The shunt resistance is determined from the inverse of the slope of the illuminated J-V curve in -0.2V to 0.0V range. In the brackets is the EQE derived J$_{sc}$ and the corresponding efficiency, which is different as the solar simulator spectra is not corrected for spectral mismatch.*

|  | Efficiency (%) | FF (%) | J$_{sc}$ (mA/cm$^2$) | V$_{OC}$ (mV) | qFLs @1sun (meV) | qFLs – eV$_{OC}$ (meV) | R$_{sh}$ (ohm-cm$^2$) |
|---|---|---|---|---|---|---|---|
| w/o treatment | 6.0 *(6.8)* | 48 | 18.8/*21.4 (EQE)* | 662 | 806 | 144 | 354 |
| AS-PDT | 7.3 *(8.4)* | 57 | 19.0/*21.8 (EQE)* | 667 | 801 | 134 | 518 |
| NaS-PDT | 7.2 *(8.3)* | 60 | 18.2/*21.0 (EQE)* | 651 | 771 | 120 | 373 |
| TU-PDT | 8.5 *(9.3)* | 61 | 20.1/*21.9 (EQE)* | 687 | 808 | 121 | 456 |

feqe

## 2.3 Metastable behavior in the electrical measurements

In the previous section, an 'S shape' was observed in the J-V curve of the untreated device [Figure 2(c)]. An 'S shape' in the J-V curve has been observed before in chalcopyrite solar cells, particularly at lower temperatures.[32-35] The presence of this 'S' shape is characteristic of a carrier transport barrier in the device and leads to attenuation of FF and $V_{OC}$.[34,36-40] In literature the presence of a highly defective layer ($p^+$) at the absorber surface is invoked as an explanation: a thin layer near the surface of the absorber which has a higher net-doping than the bulk.[35,41] The formation of this $p^+$ layer can be explained by the existence of the double vacancy defect ($V_{Cu}$ and $V_{Se}$) in $CuInSe_2$.[42] A recent study on Cu-rich $CuInSe_2$ by Elanzeery et al. supports the model of a $p^+$ layer related to Se vacancies.[21] We believe a similar defect (involving $V_S$) might also be present in $CuInS_2$ system resulting in 'S shaped' J-V curves.

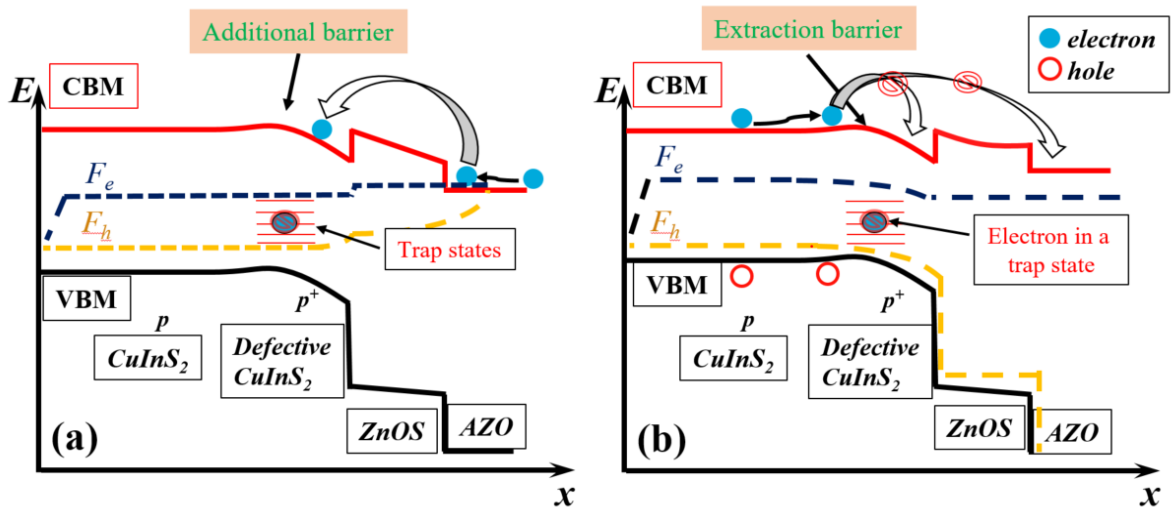

*Figure 3. Band diagram sketch of CuInS$_2$ device with a p$^+$ layer at the absorber/buffer interface (a) in the dark at moderate forward bias ~1.0V showing two barriers for injected electrons one at the Zn(O,S)/AZO and the other one in the p$^+$ layer at the front surface in the absorber (b) under illumination at V$_{OC}$ showing an extraction barrier for photogenerated minority carriers (electrons) leading to a reduced J$_{sc}$ and FF.*

A direct consequence of this p$^+$ layer at the absorber surface is a conduction band bulge near the absorber buffer interface [Figure 3(a)], which originates from the levelling of the Fermi level in the system in equilibrium. This introduces an additional barrier for injected electrons under forward bias.[36,38] As a result, a voltage drop occurs across the barrier and the forward current is reduced. Moreover, defects in this p$^+$ layer can act as minority carrier recombination centers in the SCR under illumination.[30,43] This reduces the concentration of minority carriers near the interface. Consequently, the minority quasi-Fermi level near the interface is closer to the valence band near the surface compared to the bulk. This reduces the $V_{OC}$ of the device to a value less than the measured qFLs by calibrated PL. Also, if the traps do not saturate under illumination, a higher hole concentration near the surface is maintained. This again leads to a conduction band bulge near the absorber, which leads to an energy barrier for the extraction of minority charge carriers [Figure 3(b)] and therefore to the reduction of FF and $J_{sc}$ of the device. In addition to the p$^+$ layer, any barrier such as a positive conduction band offset makes the transport even worse, diminishing the FF and $J_{sc}$ of the device even further. Therefore, in general the 'S shape' in J-V curve is more prominently visible in devices with Zn(O,S) buffer layer as compared to CdS buffer. The 'S shape' and reduction of FF and $J_{sc}$ can in many cases be mitigated by light soaking the device under open circuit conditions.[33,35,36,44-50] However, placing the device in the dark for several hours can bring the device back to the initial state i.e. with low FF and $J_{sc}$, indicating that the involved defects show metastable behavior.

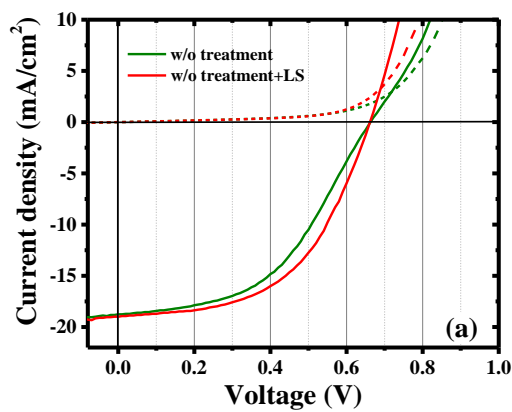
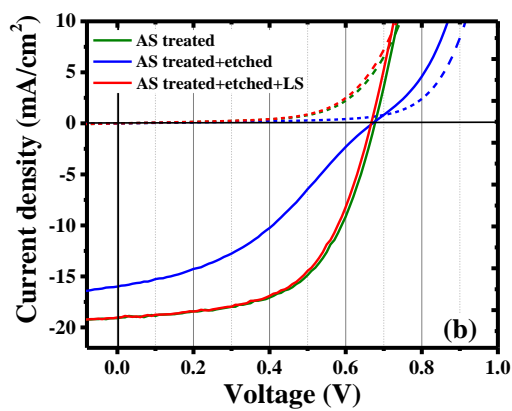

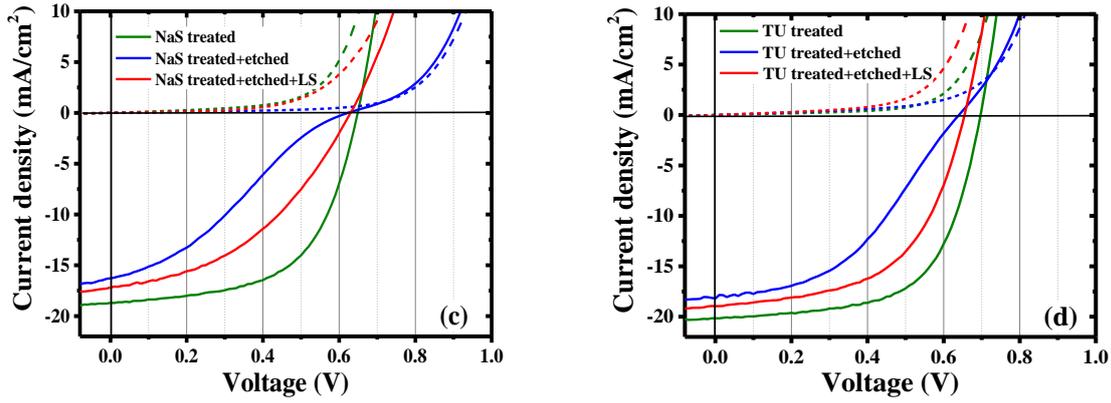

***Figure 4.*** *J-V curves of Cu-rich CuInS$_2$ device (a) without treatment and the J-V curves for three different treatments with and without etching (b) AS-PDT (c) NaS-PDT and (d) TU-PDT. An S-shape is clearly visible in the case of the untreated and of the etched devices, and disappears in all devices after light soaking (LS).*

Figure 4 shows the J-V curve of the devices that were etched after the post S-PDT along with the untreated device. In all these devices the 'S' shape is clearly observed. However, after light soaking for 30 minutes under open-circuit conditions, the 'S' shape in the J-V curve disappears. The 'S' shape appears again after keeping the device in the dark for 24 hours (not shown here). These observations suggest CuInS$_2$ also suffers from similar metastable defects (possibly related to the Cu-S double vacancy) as in the case of CuInSe$_2$. On the contrary, in the S-PDT devices which were not subject to the second KCN etching step, the 'S' shape is absent [Figure 2(c) and Figure 4]. These observations suggest the passivation of metastable defects, related to S-vacancies, in these devices. Still, even in these devices light soaking results in FF improvement [Figure 2(c) and Figure S6], and results in a maximum device efficiency of 9.0 % for the TU-PDT device (figure 5(a). It is worth noticing that all the J-V parameters of the best device improve with LS except for $V_{OC}$, which is reduced. The reduction in $V_{OC}$ is due to degradation of the device over time, something that is commonly observed in all the devices presented in this study. This degradation can be partially recovered with light soaking but not completely.

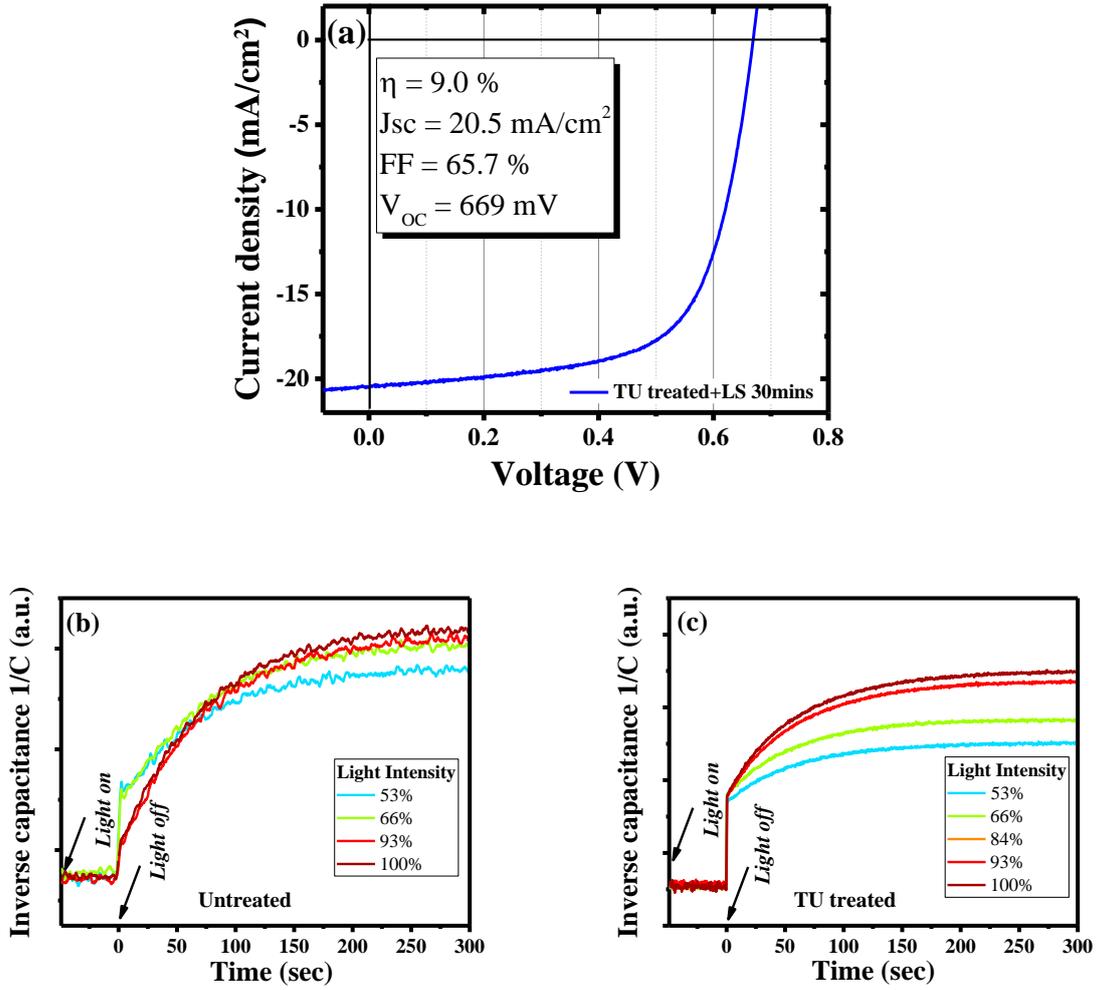

*Figure 5.* *(a) J-V curve of the light soaked TU-PDT Cu-rich CuInS$_2$ device. (b) and (c) evolution of apparent SCR width as a function of time for different illumination levels for the untreated device and TU treated device respectively. The curves are shifted to have same transient capacitance value at t = 0 seconds. The transient is measured keeping the device under illumination for 300 seconds before t=0 seconds and subsequently in the dark for 300 seconds. The device was always kept under short-circuit condition, 100% light intensity corresponds to 1 sun intensity*

In order to remove the S shape in J-V curve of the untreated or the etched device [Figure 4 and Figure S6], a considerable duration of light soaking is required, this implies the engagement of 'slow' defects. To explore the nature of the metastable defects in the devices, the time-evolution of the SCR width of the reference device and the best S-PDT device (Figure 5(b) and (c)) was analyzed. The metastable behavior after the second KCN etch step can be seen in the SI (Figure S7). The SCR width transient is measured by the inverse capacitance using the relation:

$$w(t) = \varepsilon \varepsilon_o A / C(t) \tag{2}$$

where C(t) is the measured transient capacitance as follows: first, the sample is kept under illumination with a certain intensity for 300 seconds starting from t = -300 seconds. The 'w(t)' is the transient space charge region width of the device, $\varepsilon$ is the relative dielectric permittivity of absorber taken equal to 10 as commonly used in literature,[51,52] and $\varepsilon_o$ is the dielectric permittivity of free space. It must be noted that the measured SCR width includes contribution from both absorber and buffer. However, this fact can be ignored here as we only discuss slow metastable changes of the capacitance. Throughout this illumination period, a reverse external voltage bias is applied to make sure the device always stays in short-circuit conditions (detailed information in SI), as the internal resistance of the inductance, capacitance and resistance (LCR) meter together with the short-circuit current puts the device in forward bias state. During the illumination period, the traps are occupied with photogenerated carriers, a nearly constant capacitance towards the end of the period ensures a saturation state. The illumination intensity and the applied voltage is then set to zero at t = 0 sec and the capacitance transient is measured for at least 300 seconds more. This allows the device capacitance to reach a constant value (after de-trapping of carriers) indicating the device is in a new certain quasi-steady state.

Figure 5(b) and (c) shows the SCR width transients in the dark after illumination. Transients under illumination are shown in Figure S7 of the supplementary information. The device capacitance is higher and thus the effective SCR width lower under illumination due to the additional contribution of light-generated charge carriers. We are mostly interested in the SCR width change between the illuminated state and the dark state at t = 0 sec. The device capacitance transient was measured with lowest illumination intensity first followed by higher illumination intensity. After each measurement, the device is in a quasi-steady state, which is different from the previous steady state. Bringing the device back to the completely relaxed state is extremely time-consuming. Therefore, all the curves have been shifted to the same SCR width value at t = 0 seconds to allow a better comparison, the unshifted curves can be found in SI. For the untreated device, at t = 0 seconds, the SCR width increases abruptly [Figure 5(b)], as the excess of light generated carriers recombines. This fast increase in SCR width in this device decreases with the increase in illumination intensity. This is a consequence of trapping of excess charge carriers in the deep recombination centers, which release these charge carriers slowly. The gradual (slow) increase of the SCR width with time is due to the release of charge carries from the deep defects. It is interesting to note, the slow change behaves opposite to the quick jump: less slow change at lower intensities, because less photogenerated carriers are trapped. The transients of the TU-PDT device [Figure 5(c)] show a different behavior: the abrupt change in the SCR width upon switching off light at t = 0 sec is independent of the illumination intensity. This can be understood as a direct consequence of the passivation of the 'slow' defects in the TU-PDT device. For the untreated device, these defects trap charge carriers, more so with higher illumination. After switching off the illumination, the defects release the charge carriers slowly. Under 1 sun illumination, the charge response after illumination is almost entirely given by the slow defects, there is only a very small jump in SCR width at t = 0 seconds, which can be attributed to free carriers, and almost the entire transient back to the dark state is due to slow defects. On the contrary, (partial) passivation of these defects after TU-PDT results in much less carrier trapping. The quick free-carrier response is always visible. However, the fact that the jump in SCR width remains the same and does not increase with illumination suggests that some photogenerated carriers are still trapped. The slow transient following the first jump also indicates this. The magnitude of this slow response increases with illumination, indicating more carriers trapped in slow metastable defects with higher illumination. This shows that some of these defects remain after the treatment. To complete the series, the SCR width transient measurements for the post S-PDT KCN etched device was also probed. Like the J-V measurements, this device shows a transient response similar to the untreated device [Figure S7(c)], indicating the removal of the beneficial impact of TU treatment. The KCN etching is already known to preferential removal Cu and Se atoms from the Cu-rich CuInSe$_2$ lattice and thus forming ($V_{Cu}+V_S$) divacancy defect complex.[53] We hypothesize a similar mechanism also applies to Cu-rich CuInS$_2$ absorbers.

In summary: the response of the untreated device is dominated by slow defects, this response increases with higher illumination intensity, whereas the treated device shows much less response of slow defects accompanied with a free carrier response. Thus, both J-V and capacitance transient measurements show the effectiveness of S-PDT, especially TU-PDT, in the passivation of near surface defects. In addition, it also shows that these slow defects have characteristics, which are usually associated with metastable defects.

## 2.4 Interface recombination analysis

The low $V_{OC}$ compared to the bandgap has been attributed to interface recombination in Cu-rich chalcopyrite solar cell.[9,21,54,55] In addition, the large deficit between the quasi-Fermi level splitting and the $V_{OC}$ (see table 1) is a result of recombination at or near the interface.[55,56] Temperature-dependent current-voltage (JVT) analysis can help identify the dominant recombination pathway in the device:[13]

$$J_0 = J_{00} \exp\left(\frac{-E_a}{AkT}\right) \quad (3)$$

Using this together with equation 1:  $\quad V_{oc} = \dfrac{E_a}{q} - \dfrac{AkT}{q} \ln\left(\dfrac{J_{00}}{J_{sc}}\right) \quad (4)$

where $R_{sh}$ is ignored and $J_{oo}$ is a weakly temperature-dependent pre-factor and $E_a$ is the activation energy of the dominant recombination process. From equation 4, a linear temperature-dependent $V_{OC}$ extrapolation to 0 K yields the $E_a$ of the dominant recombination process, assuming the diode factor A and the $J_{sc}$ to be constant (at least at moderately high temperatures i.e. 220-300 K).

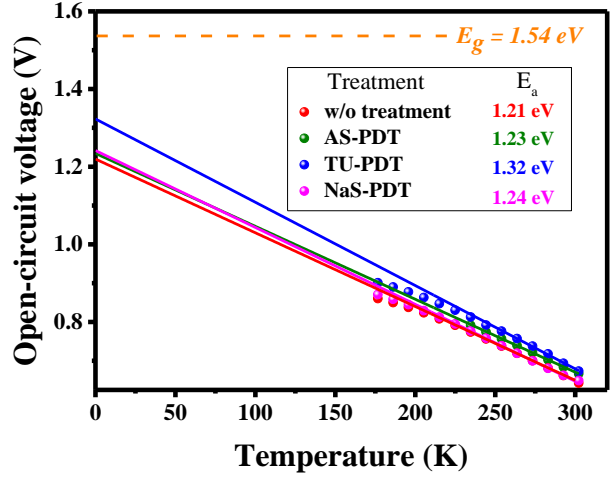

*Figure 6. Open circuit-voltage vs temperature of Cu-rich CuInS$_2$ devices without any treatment, with AS-PDT, NaS-PDT and TU-PDT. The band gap was determined from low energy inflection point in the external quantum efficiency spectrum.*

Figure 6 shows the temperature dependence of the $V_{OC}$ of the device without any treatment and devices with AS, NaS and TU-PDT, the legend shows the activation energies. This is obtained by linear extrapolation of the high-temperature $V_{OC}$ to 0 K. The bandgap of the devices was unaffected by treatments and was determined by the inflection point in the external quantum efficiency spectra (Figure 2(d)). For all devices the activation energy $E_a$ is considerably lower than the bandgap energy $E_g$ of the absorber. These results indicate major recombination at the absorber buffer interface in all devices. Nonetheless, the TU-PDT does improve the activation energy of the devices, in agreement with the device performance trends (see table 1). This shows a TU-PDT can particularly help reduce interface recombination in the final device.

## 3. Discussion: Impact of S-salt in S-PDT

Throughout this work, the three S-PDTs have shown a different impact on device performance, with TU-PDT being the most effective among the three S-PDTs. The three S-PDTs have the same concentration of sulfur atoms in the solutions, but are different from each other with respect to the sulfur source, exact sulfur species in solution, solution pH, and the cations, as shown in table 2. The PDTs AS and NaS share the same sulfur source, namely the anion $S^{2-}$, whilst for TU, the origin of the sulfur is the S atom covalently bound to a carbon atom. We discuss the AS and NaS PDTs first. In an aqueous solution, AS ($(NH_4)_2S$) dissociates into $NH_4^+$ and $S^{2-}$ and NaS ($Na_2S$) dissociates into $Na^+$ and $S^{2-}$ ions.[57,58] We assume that the $S^{2-}$ ion diffuses to the surface of the absorber and reacts to fill any anion vacancies. However, the exact sulfur species present in solution depends on the pH, with the more basic NaS PDT having a higher proportion of $S^{2-}$ ions to $HS^-$ ions as compared to the AS PDT.[59] The pH of the PDT solution might further impact the absorber by in-parting a different surface charge since this depends on the presence of the potential determining ion $HS^-$,[60] although in the subsequent solution based deposition of Zn(O,S) this difference may be equalized again. One further detail is the presence of the cation species in the solution. The $Na_2S$ solution contains $Na^+$ ions which we hypothesize that they can aggregate at grain boundaries or even go on to Cu sites in the absorber grains themselves since sodium is very mobile.[61] Both of these possibilities could have led to passivation of recombination centers and perhaps higher $V_{OC}$ in the device. Unfortunately though, this solution also led to delamination of the absorber from the Mo substrate during our experiments, and this accounts for the slightly degraded device performance compared to AS.

Table 2. Summary of sulfur-post deposition treatment conditions and chemical species present in each solution

| PDT | Sulfur source | Cations | Anions | PDT pH |
|---|---|---|---|---|
| *AS* | 0.4M $S^{2-}$ | 0.8M $NH_4^+$ | - | 9.1 |
| *NaS* | 0.4M $S^{2-}$ | 2M $NH_4^+$, 0.8M $Na^+$ | 2M $OH^-$ | 13.2 |
| *TU* | 0.4 M S=C(NH$_2$)$_2$ | 2M NH$_4^+$ | 2M OH$^-$ | 11.6 |

The case for TU is more complicated. Previously it was believed that TU releases sulfur in basic conditions *via* HS$^-$ to form S$^{2-}$.[62,63] However, a new study suggests that TU only releases sulfur when it is directly complexed to a metal cation.[64] Assuming this is correct, the TU PDT must act differently to the AS and NaS-PDTs. Whilst AS and NaS-PDTs rely on the active sulfur species diffusing to the surface of the absorber layer, in the TU-PDT the CH$_4$N$_2$S molecule must diffuse and then physically adsorb first. Once adsorbed it may remain physisorbed or react with the surface of the semiconductor. If the TU physisorbs at the surface, it would lead to a thin protective layer, which may act as a physical protective barrier during the buffer layer deposition, maintaining the absorber quality. Surface X-ray photoelectron spectroscopy (XPS) analysis of the absorber treated with TU supports this argument of reaction of TU with the surface along with the presence of an additional layer linked to TU. Compared to untreated and AS-PDT absorber, XPS data of the TU-PDT absorber displays the presence of a second S compound, compatible with an organic sulfur species (likely relating to TU) in addition to CuInS$_2$ (see discussion in SI). This might explain why the qFLs does not change after depositing a buffer layer on the TU treated absorber and at the same time provides passivation of surface defects. Further, the fact that the device with TU-PDT device has different interference pattern in EQE also suggests the modification of the absorber surface.

To summarize, the different effects of the various PDTs can be due to different S-species in solution, which are expected to show different absorption behavior and reaction mechanism behavior. Further investigation of the PDTs is required to understand the reaction mechanism better. However, we would like to stress, that a simple treatment with an S containing solution does improve the interface of the solar cells.

## 4. Conclusion

Interface recombination is one of the main factors for the low efficiency of Cu-rich CuInS$_2$. Using a simple sulfur solution immersion technique we were able to show partial passivation of these defects by using thiourea as a sulfur source. We probed the optoelectrical properties of Cu-rich CuInS$_2$ before and after the treatment. After buffer layer capping, the qFLs decreased in each case except for the absorber with thiourea PDT which also translated into better device performance. The current-voltage characteristics of these devices showed improved FF and efficiency with each of these treatments. However, not all the treatments were equivalent, the thiourea PDT clearly showed superior passivation ability, as indicated by the highest open circuit voltage. On the other hand, sodium sulfide PDT did decrease the open circuit voltage. Thus, the sulfur source in the treatment solution can also have an adverse impact and hence must be chosen carefully. The best device performance was obtained by using thiourea as the sulfur source as it improves the interface without any adverse effect on the final device properties. This was manifested in our J-V and JVT measurements.

In the untreated device and in devices, etched again after the S-PDT, metastabilities were observed with the help of current-voltage and slow capacitance transient measurements, and were related to a sulfur vacancy related defect. This defect can be partially passivated by sulfur treatment, thiourea in particular, as demonstrated by our capacitance transient measurements, and can be again created at the surface by KCN etching.

The facile solution based treatment demonstrated in this study is also applicable to other thin film solar cells, in particular chalcogenide based ones. It can be expected that similar treatments can be developed to mitigate interface recombination and improve open-circuit voltage.

## 5. Experimental Methods

*Absorber preparation:* For the experiments, 2-stage absorbers were grown at 590 °C on molybdenum sputtered onto soda-lime glass (SLG) in our standard process [7] with Cu-rich elemental composition ([Cu]/[In] ~1.7 and [S]/([Cu]+[In]) ~ 0.98, as measured by energy-dispersive X-ray spectroscopy (EDX)). The 2-stage process was preferred as it allows the formation of a compact layer with a smooth surface.[7] These are important requirements to reduce shunting paths in the final device. Following the absorber growth, a 10 % potassium cyanide (KCN) etching for five minutes was performed to remove the $Cu_{2-x}S$ secondary phase.[65]

*Post deposition treatment and device preparation:* After KCN etching, the absorbers were subjected to the sulfur treatment followed by Zn(O,S) buffer layer deposition. One absorber was not treated and directly covered with Zn(O,S) buffer layer, in order to have a reference device. The details for Zn(O,S) buffer layer deposition can be found in the supplementary information and are based on a recipe by Hubert et al..[66] For the S-PDT: three separate aqueous solutions consisting of $(NH_4)_2S_x$ (0.4M), $Na_2S$ (0.4M) in $NH_4OH$ (2M) and $CH_4N_2S$ (0.4M) in $NH_4OH$ (2M), respectively, were freshly prepared in deionized water (18.2 M-ohm resistivity) immediately before starting the treatment. Each of these solutions was heated to 80 °C on a hot plate. Then six freshly KCN etched $CuInS_2$ absorbers were immersed in each of the three different solutions (two absorbers in one solution) for 10 minutes and afterward rinsed with DI water. After the sulfur treatment, one absorber from each treatment was again subject to 5% KCN etching for 30 seconds. The aim of this etching was to remove the passivating effect of the S-PDT (if any). All of these absorbers (3 treated, 3 treated and afterwards etched, and 1 untreated) were processed with a Zn(O,S) buffer followed by a sputtered i-ZnO (80nm) and Al:ZnO (380nm) window layer. Figure 1 shows the entire schematic of the previously described process. On top of the window layer, a Ni-Al dot was evaporated using an e-beam for electrical contact. Finally, the devices with SLG/Mo/CuInS$_2$/Zn(O,S)/i-ZnO/AZO architecture with an area of around 0.2 cm$^2$ were delineated using mechanical scribing. For quasi-Fermi level splitting measurements, 1 absorber from the same run was cut into small pieces. These small absorber pieces were then S-treated exactly in the same manner as the absorbers for solar cells, and then coated with Zn(O,S).

*Characterization methods:* The elemental composition was measured using energy dispersive X-ray spectroscopy in scanning electron microscope (SEM) with an operating voltage of 20 kV and for cross-section images operating voltage of 7 kV was used. The current-voltage (J-V) properties of the device were investigated using an AAA-Standard solar simulator with a Xenon short-arc lamp, calibrated by a Si reference cell, with an IV source-measure-unit. The external quantum efficiency measurements were performed using grating monochromator setup and a chopper, halogen and xenon lamps as light source. The current is measured using a lock-in amplifier and the intensity of the light by calibrated reference diodes. Four-point admittance and temperature-dependent J-V measurements were performed by mounting the samples in a closed-cycle cryostat chamber, with a base pressure below 4x10$^{-3}$ mbar. For measuring the temperature-dependent J-V, a cold mirror halogen lamp was used as a source for illumination. The height of the lamp from the sample was adjusted to an equivalent intensity of 100 mW/cm$^2$, by controlling for a J$_{sc}$ equal to the one measured under the solar simulator. Precise measurement of the sample temperature was made by gluing a Si-diode sensor onto an identical glass substrate and placing it just beside the solar cell. Capacitance transients were recorded using an LCR meter with a controlled small-signal ac voltage pulse of 30 mV rms at a frequency of 10 kHz. Photoluminescence (PL) measurements were carried out in a home-built system using a CW 663 nm diode laser as an excitation source. For the determination of the quasi-Fermi level splitting, intensity and spectral corrections were applied to the raw data, more details can be found in these reports.[7,28]

**Supporting Information**

Supporting Information is available from the Wiley Online Library or from the author.

**Acknowledgment**

This research was funded in whole, or in part, by the Luxembourg National Research Fund (FNR), grant reference [PRIDE 15/10935404]. For the purpose of open access, the author has applied a Creative Commons Attributions 4.0 International (CC BY 4.0) license to any Author Accepted Manuscript version arising from this submission. We also thank and acknowledge Prof. Małgorzata Igalson for her suggestions, feedback on the manuscript.

# Supporting Information

**Passivating Surface Defects and Reducing Interface Recombination in Solar Cells by a Facile Solution Treatment**

*Mohit Sood, Alberto Lomuscio, Florian Werner, Aleksandra Nikolaeva, Phillip J. Dale, Michele Melchiorre , Jérôme Guillot, Daniel Abou-Ras, Susanne Siebentritt\**

M. Sood, Dr. A. Lomuscio, Dr. F. Werner, Prof. P.J. Dale, Dr. M. Melchiorre, Prof. S. Siebentritt

Department of Physics and Materials Science, University of Luxembourg, Belvaux, L-4422, Luxembourg

E-mail: susanne.siebentritt@uni.lu

*Dr. J. Guillot*

Luxembourg Institute of Science and Technology, Belvaux, L-4422, Luxembourg

A. Nikolaeva Dr. D. Abou-Ras

Helmholtz-Zentrum Berlin für Materialien und Energie GmbH, Hahn-Meitner-Platz 1, 14109 Berlin, Germany

**Influence of thiourea concentration in chemical bath and buffer layer deposition**

In our experiments, both the CdS and the Zn(O,S) buffer layers were deposited onto the absorbers by the well-established chemical bath deposition (CBD) technique.

Concerning the CdS buffer layer, we fabricated two different devices by varying the TU concentration. For the first device, CdS-1, we used a standard TU concentration (50 mM), whereas for the second one, CdS-2, we tripled it (150 mM). Generally, 2 mM of cadmium sulfate hydrate (Alfa Aesar, CAS 15244-35-6) were dissolved in a 1.5 M aqueous ammonium hydroxide solution (Honeywell, CAS 1336-21-6) at 67 °C. After 3 minutes, TU (Sigma

Aldrich, CAS 62-56-6) was added to the solution. The reaction was stopped after the solution turned from a transparent to a turbid yellowish color, meaning roughly 40-50 nm of CdS buffer layer were deposited. Since it is well known that the band alignment between the CdS buffer layer and the $CuInS_2$ absorber is unfavorable,[8,54] an additional device with Zn(O,S) buffer layer was also fabricated as follow. In a double-jacketed reactor, 0.1 M of zinc sulfate heptahydrate (Sigma Aldrich, CAS 7446-20-0) were dissolved in a 2 M aqueous ammonium hydroxide solution (Honeywell, CAS 1336-21-6) at 84 °C, then 0.4 M of TU (Sigma-Aldrich) were added to the solution and the Zn(O,S) deposition started. The reactor was kept constantly at 84 °C during the whole CBD process. To obtain the optimal 50 nm Zn(O,S) buffer layer thickness, two runs with the same recipe were needed and performed on the same absorber. The Zn(O,S) CBD process was taken and adapted from the work of N. Naghavi et al.[66]. It is worth highlighting that in the case of the Zn(O,S) buffer layer CBD deposition, TU was 8 times more concentrated than our standard CdS buffer layer methodology.

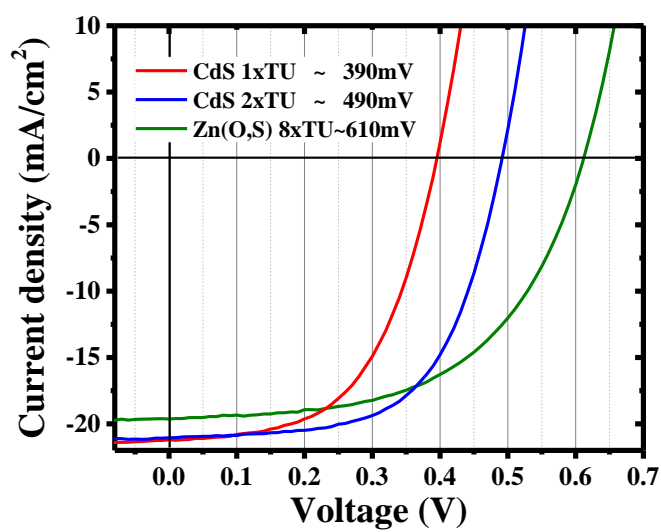

*Figure S1. JV characteristics of $CuInS_2$ device with CdS buffer 1xTU, 2xTU and Zn(O,S) buffer with 8xTU in chemical bath solutions.*

The JV characteristics of the devices with the different buffer layer, CdS1 (1xTU), CdS2 (3xTU), and Zn(O,S) (8xTU) are shown in Figure S1. There is a clear improvement in device performance, especially the open-circuit voltage with an increase of TU concentration in the chemical bath. This suggests a definitive effect of TU concentration in the chemical bath solution on the open-circuit voltage of the devices. Note, the higher $J_{sc}$ in devices with CdS buffer is because our baseline ZnO and Al:ZnO is optimized for the device with CdS buffer not Zn(O,S), and hence we have higher reflective loss in device with Zn(O,S).

It worth mentioning that in $CuInS_2$ devices with CdS buffer layer the 'S shape' is not observed. This is because the additional barrier (other than $p^+$ layer) due to a positive conduction band offset is not present in these device.[16,18] However, we do notice an 'S shape' in some of our $Cu(In,Ga)S_2$ devices prepared with CdS buffer layer. This is a result of an increase in front Ga grading (*i.e.* increase in Ga atomic concentration towards the surface), which leads to an increase the conduction band towards the surface. Thus providing an additional barrier to minority carriers together with the $p^+$ layer, similar to a positive conduction band offset at the absorber buffer interface.

**Calibrated photoluminescence measurement and quasi-Fermi level splitting determination**

The calibrated photoluminescence measurements to extract the qFLs have been performed under an equivalent illumination of five suns to ease the spectra acquisitions, to allow for faster and more reliable measurement, in spite of the quite low radiative efficiency of these absorbers. Then, those values have been corrected for one sun illumination, as listed in table 1 in the main part of the manuscript. The correction is based under the assumption that the optical diode factor *k* is unity, which is defined as:

$$I_{PL} \propto \phi^k \tag{5}$$

with $I_{PL}$ and $\phi$ being respectively photoluminescence intensity and excitation density.

Under this assumption, the external radiation efficiency (ERE) is constant as well, as it is defined as the ratio between the integrated PL photon flux density and the incident photon flux density. Because of the low luminescence efficiency, we could not determine *k* for these samples.

The qFLs is related to the generation under illumination (G) and recombination in thermal equilibrium ($R_0$) by the following relationship [67]:

$$qFls = k_B T * \ln\left[\left(\frac{G}{R_o}\right) * ERE\right] \quad (6)$$

with $k_B T$ the thermal energy. With ERE constant at different excitations (in the present case at 1 and 5 suns), the qFLs at 1 sun is thus determined:

$$qFls_{1sun} = qFls_{5suns} - k_B T * \ln\left(\frac{G_{5suns}}{G_{1sun}}\right) = qFls_{5suns} - k_B T * \ln(5) \quad (7)$$

$$\Rightarrow qFls_{1sun} = qFls_{5suns} - 40 meV \quad (8)$$

Experimentally we often find k>1,[55] i.e. the ERE is higher at higher excitation intensity. In this case the qFLs at 1 sun would be even smaller, since equation (8) overestimates the qFLs at 1 sun. But the trends we discuss between samples would remain the same. Thus, we consider a worst case scenario when it comes to determining the additional Voc loss due to interface recombination.

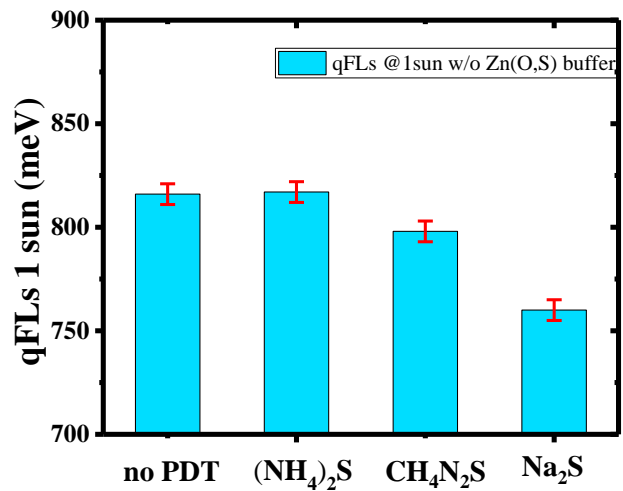

*Figure S2. Quasi-Fermi level splitting values of Cu-rich CuInS$_2$ absorber treated with AS-PDT, NaS-PDT and TU-PDT, and without any PDT without Zn(O,S) buffer calculated for 1 sun illumination from equation (8) using 5 sun calibrated photoluminescence measurements.*

**Experimental details of cathodoluminescence:**

The scanning electron micrographs and cathodoluminescence (CL) hyperspectral images were recorded using a Zeiss Merlin scanning electron microscope equipped with a DELMIC CL system at 10 keV beam energy and at beam currents of 500-700 pA. Figure S3 shows the CL images obtained on the cross-section of untreated and TU-PDT sample. Both samples show rather low CL intensity. No difference in grain boundary activity was observed.

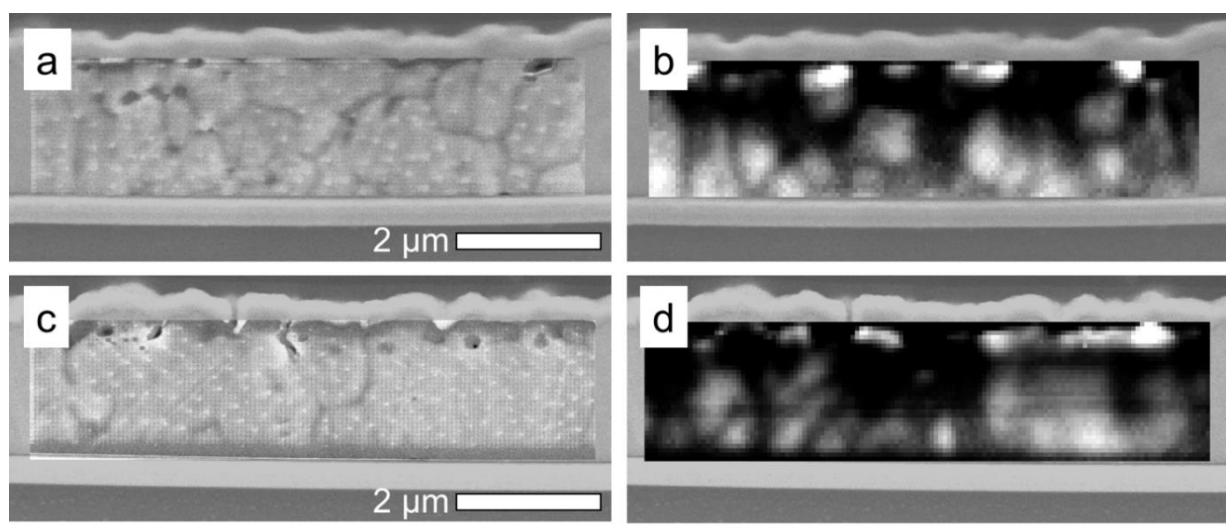

*Figure S3. SEM images (a, c) and panchromatic CL images (b,d) acquired on cross-section specimens from CuInS2 solar cells with (a,b) and without TU treatment (c,d).*

*Table S1. Device characteristic of CuInS$_2$ with different treatment and treatment plus etching.*

|  | η (%) | V$_{oc}$ (mV) | J$_{sc}$ (mA/cm$^2$) | FF (%) | R$_{sh}$ (Ω-cm$^2$) dark | R$_{sh}$ (Ω-cm$^2$) light |
|---|---|---|---|---|---|---|
| w/o treatment | 7.0 | 648 | 18.7 | 58 | 1204 | 354 |
| AS-PDT | 7.3 | 667 | 19.0 | 57 | 1310 | 518 |
| AS-PDT+etch | 7.4 | 679 | 19.6 | 56 | 2164 | 241 |
| NaS-PDT | 7.2 | 651 | 18.2 | 60 | 714 | 373 |
| NaS-PDT+etch | 5.2 | 644 | 17.4 | 47 | 2049 | 171 |
| **TU-PDT** | **8.5** | **687** | **20.1** | **61** | **1250** | **456** |
| TU-PDT+etch | 6 | 661 | 17.5 | 52 | 289 | 775 |

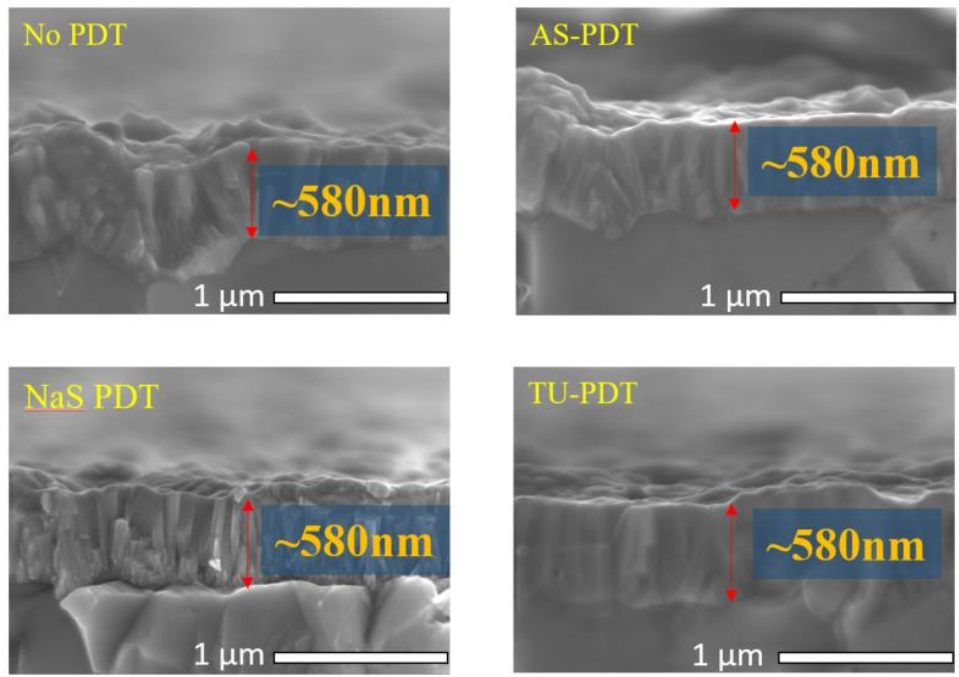

*Figure S4. SEM cross-section images of the CuInS$_2$ devices without any treatment and with AS-PDT, NaS-PDT and TU-PDT.*

**J-V fit**

J-V fit was attempted to investigate the effectiveness of S-PDT in reducing recombination in the device and is shown in Figure S5. The fitting was done using the IV orthogonal distance regression fit routine [68] by a python script. Results of the J-V fit under are summarized in table S2 for all devices. The V$_{oc}$ reported in the table is obtained using the 1-diode model from the relationship (1). [31]

As evident from equation 1, the $V_{oc}$ improves with the increase of both $J_{sc}$ and $R_{sh}$, and with the decrease in reverse saturation current. For the state of the art solar cells, the $R_{sh}$ is usually high enough such that the ratio of $V_{OC}$ to $R_{sh}$ is negligible compared to $J_{sc}$.[46] In such cases, $R_{sh}$ does not affect the $V_{oc}$ and can be neglected in the above expression. However, for solar cells, where the $R_{sh}$ is low, $V_{oc}/R_{sh}$ becomes non-negligible with respect to $J_{sc}$. The low $R_{sh}$ provides an alternative path for the light generated current reducing the $V_{oc}$ of the solar cell. Even though in our devices the $R_{sh}$ is low still the impact of it on $V_{oc}$ will be just a few mVs. From Figure S3 (a) it is clear that the fit in the range 0.0_1.0 V does not converge for the untreated device, hence a fit was also attempted in the range, dominated by the diode behavior, i.e. the straight part in the semi-log plot between ~0.4 and ~0.5eV. The results are reported in table S2. Either way, the $V_{oc}$ values obtained from fitting values do not agree with experimentally measured $V_{oc}$ values of this device. This might be because this device does not show a well-defined diodic J-V behavior. Among the S-PDT devices, the $V_{oc}$ values obtained from the fit parameters for the light curve are more in accordance with the experimental $V_{oc}$ (at least the trends are the same). Hence, it is possible to make comparisons among these devices even though the ideality factors are around 3. The obtained $J_o$ (both from fit in 0.0_1.0V and 0.5_0.7V region) for TU-PDT and NaS-PDT indicates better passivation of recombination centers in comparison to the AS-PDT which has a higher $J_o$ (see table S2). This follows the qFLs values after buffer, which remains unchanged after both TU-PDT and NaS-PDT, whereas, decreases for AS-PDT (see Figure 2). Among the three treatments studied, solely TU-PDT results in an improvement in both: a higher $R_{sh}$ and a lower $J_o$. Thus, improvement in both $R_{sh}$ and $J_o$ leads to the device with highest FF, $V_{oc}$ and efficiency.

We would like to point out that all these devices showed performance degradation with time. This degradation can be partially recovered with light soaking but not completely.

***Table S2.*** *J-V fit results of CuInS$_2$ device with different treatments. Two different fitting ranges were tried: one to fit the whole diode curve in forward bias and the other to just fit the diode part. The shunt resistance reported here is obtained from inverse slope of J-V curve in the range -0.2 to 0.0V. The V$_{oc}$ values are obtained by inserting the obtained fit values into equation 1.*

| Treatment | Fit range | A | | J$_o$ (µA/cm$^2$) | | R$_{sh}$ (kΩ) | | Calculated V$_{oc}$ (mV) | |
|---|---|---|---|---|---|---|---|---|---|
| | | Dark | Light | Dark | Light | Dark | Light | Dark parameters | Light parameters |
| w/o PDT | 0.4V-0.7V | 5.0 | 5.0 | 8.2 | 732 | 1.2 | 0.35 | 993 | 419 |
| | 0.0V-1.0V | 3.5 | 5.0 | 0.8 | 0.8 | 1.2 | 0.35 | 899 | 900 |
| AS PDT | 0.5V-0.7V | 2.9 | 3.0 | 0.6 | 3.7 | 1.3 | 0.52 | 772 | 652 |
| | 0.0V-1.0V | 3.0 | 3.6 | 0.9 | 14.2 | 1.3 | 0.52 | 769 | 664 |
| NaS PDT | 0.5V-0.7V | 2.4 | 2.2 | 0.36 | 0.23 | 0.7 | 0.37 | 676 | 650 |
| | 0.0V-1.0V | 2.6 | 2.8 | 0.65 | 2.2 | 0.7 | 0.37 | 682 | 649 |
| TU PDT | 0.5V-0.7V | 2.1 | 2.2 | 0.03 | 0.16 | 1.2 | 0.45 | 723 | 666 |
| | 0.0V-1.0V | 2.5 | 3.1 | 0.2 | 3.6 | 1.2 | 0.45 | 749 | 680 |

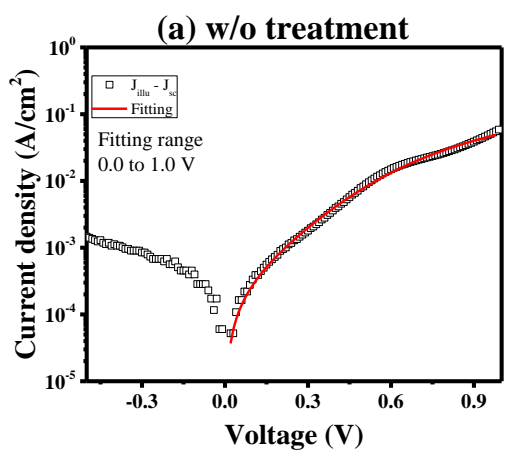

(a) w/o treatment

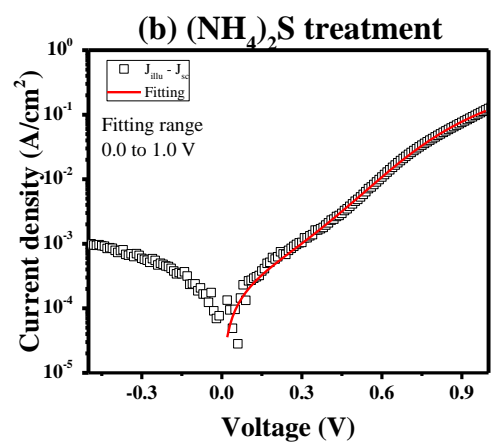

(b) (NH$_4$)$_2$S treatment

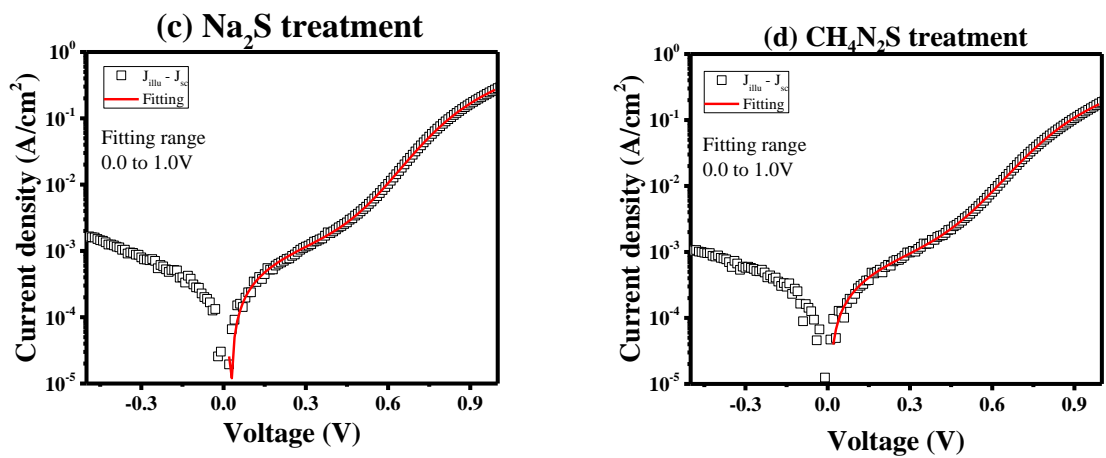

*Figure S5*. Fitting of illuminated JV curve in the voltage region 0.0_1.0 V shifted by the short-circuit current of CuInS$_2$ device (a) w/o treatment (b) AS-PDT (c) NaS-PDT and (d) TU-PDT. The curves shown here are for devices without light soaking. The results of fitting i.e. the values of A, J$_o$, are reported in table S2.

*Table S2*. *Characteristic of Cu-rich CuInS$_2$ device treated with AS-PDT and NaS-PDT with and without light soaking*

|  | Efficiency (%) | V$_{oc}$ (mV) | J$_{sc}$ (mA/cm$^2$) | FF (%) |
|---|---|---|---|---|
| *w/o treatment* w/o light soaking | 6.0 | 662 | 18.7 | 48 |
| ***w/o treatment* with light soaking** | **6.6** | **660** | **19.0** | **53** |
| *AS-PST* w/o light soaking | 6.9 | 662 | 18.1 | 57 |
| ***AS-PDT* with light soaking** | **7.9** | **661** | **18.6** | **64** |
| *NaS-PDT* w/o light soaking | 7.2 | 651 | 18.3 | 60 |
| ***NaS-PDT* with light soaking** | **7.8** | **649** | **19.5** | **61** |

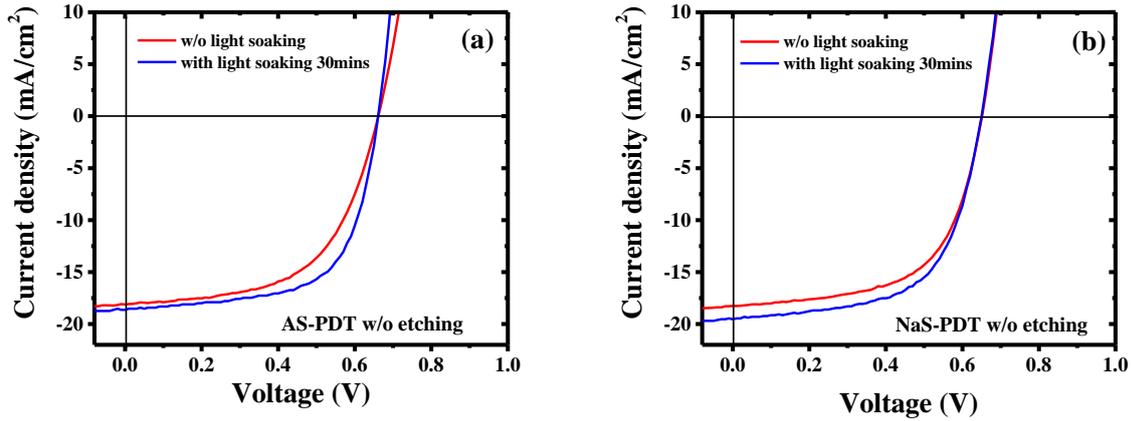

*Figure S6.* JV characteristics of CuInS$_2$ device before and after light soaking (a) AS-PDT (b) NaS-PDT.

**Transient capacitance measurements**

The procedure to measure the capacitance transients is as follows: first, the sample is kept under illumination with certain intensity for 300 seconds starting from t = -300 seconds. Since the LCR meter has an internal resistance of about 100 ohm, this resistance under illumination puts the device in a certain forward biased state. Hence to keep the device under short-circuit conditions a reverse bias voltage is applied to compensate for the photo-voltage from t = -300 seconds to 0 seconds i.e. for the whole illumination period. This was done by measuring the DC voltage generated across the device due illumination, using the LCR meter. Thereafter, a voltage exactly opposite to this measured voltage is applied when the device is under illumination to keep the device under short-circuit condition. During the entire measurement procedure the voltage is monitored to make sure the device is always under short-circuit conditions. After this first step of 300 seconds, the illumination intensity is then set to zero at t=0 sec and the capacitance transient is measured for at least 300 seconds more. Note, no bias was applied on the sample during this second step i.e. for t ≥ 0 seconds. Figure S7 shows the evolution of space charge region width with time in three samples: untreated, TU-PDT and TU-PDT followed by KCN etching.

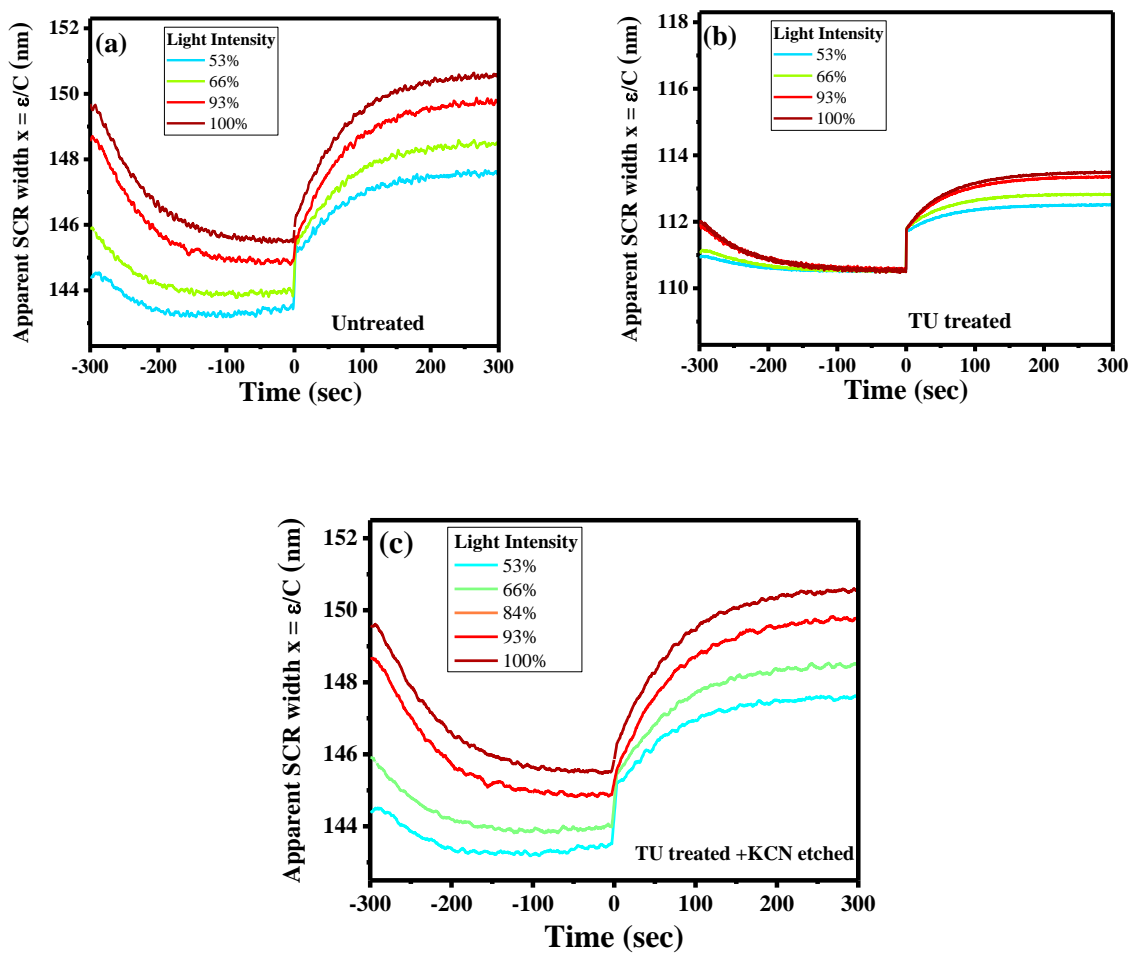

*Figure S7. Evolution of apparent SCR width as a function of time for (a) untreated sample (b) TU-PDT treated (c) TU-PDT +KCN etched device; measurements were done keeping the device under illumination for 300seconds and subsequently under dark for 300seconds. Always keeping the device in short-circuit condition: 100% light intensity is equivalent to 1 sun intensity*

## Surface analysis by X-ray photoelectron spectroscopy

To check for the chemical impact of S-PDT on the absorber surface, X-ray photoelectron spectroscopy (XPS) was performed on the samples. For this, four pieces of 10% KCN etched Cu-rich $CuInS_2$ absorbers from the same absorber deposition run were used. First piece was left untreated (ref), the second was treated AS-PDT (sample '1'), third one with TU-PDT (sample '2') and the last one with NaS-PDT, which delaminated and was not analyzed. All the three remaining pieces were then transferred with water layer (comes from rinsing the absorber with DI water after KCN and the S-PDT) on top into the glove box to avoid air exposure. From the glove box they were then transferred via a N$_2$ filled cell into the XPS chamber for analysis. The entire procedure was designed to ensure the minimum air exposure. XPS experiments were carried out using a Kratos Axis Ultra DLD instrument equipped with a monochromatic Al Kα source (1486.6 eV) working at 150W. The base pressure during the analyses was better than 5.10$^{-9}$ mbar. The narrow scans, for elemental quantification and chemical states investigations, were recorded with an energy resolution of 0.6 eV. The samples were sputtered with monoatomic Ar$^+$ ions of low energy (500 V) to limit the preferential sputtering effects, for 180 s to remove the surface contaminants and for 1080 s to access the deeper composition. The data were processed with the CasaXPS software (v2.3.22) and the curve fitting obtained with 70% Gaussian – 30% Lorentzian lineshapes.

Figure S8 (a) shows the S 2p bulk spectra of the above-mentioned three samples. For TU-PDT absorber, proper fitting of 'S' spectrum required fitting with two doublets. The two peaks results from the S 2p$_{3/2}$ and S 2p$_{1/2}$ spin-orbit split: the first doublet correspond to CuInS$_2$, which is present in spectra of all samples, and an additional doublet with peaks at 164.2-165.1 eV to account for bump at higher binding energy that could be signature of C-S-C. [69-71] Simultaneously TU-PDT sample also has significant amount of N present on the surface which is absent in the other two samples. Analysis of N 1s bulk spectra shows peak at 399.3 eV, which corresponds to C-NH2. [72] These results suggest the presence of an additional phase of S and an organic component containing C-NH2. At the surface, from the elemental quantification amino groups are present in 3.5 at% concentration, which implies the presence of 1.75 at% of TU. Further, from the S2p spectrum, the organic component (S2p-2) represents only 1.1 at% of the total composition (6.6 % of the total sulfur area, and S being 16 at% of the total composition). This suggests that not all the amino groups are present in form of TU or in the organic sulfur phase. Rather a portion of TU has partially reacted leaving amino group behind at the surface. Thus, the results indicate that a part of TU has reacted with the surface (physisorption of TU particular S) while the other part of TU is still present at surface in the form of fragments. Additionally, table S3 represents the [Cu]/[In] ratio obtained using Cu 3p, Cu 2p, In 3d and In 4d lines. With the analysis depth for the different lines (in a CuInS2 matrix): Cu 3p = 7.9 nm; Cu 2p = 4.2 nm; In 3d = 6.25 nm; and In 4d; 8.2nm (calculated from the TTP2M formula). [73] Unlike the other two samples the [Cu]/[In] ratio changes dramatically with the use of different element lines. Thus, indicating the presence of thin layer on top, as this over layer influences the analysis depth of each element. Therefore, XPS analysis concludes the formation of an organic over layer on the absorber surface after TU-PDT.

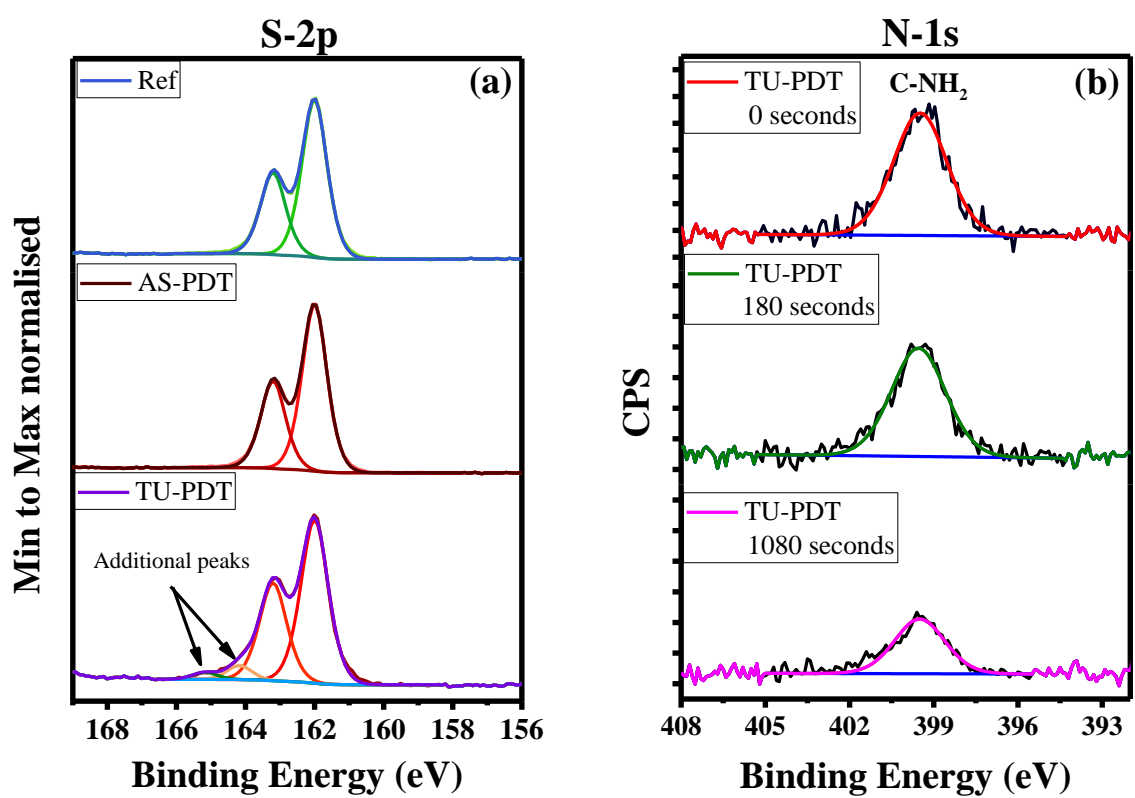

*Figure S8.* (a) S-2p core level spectrum acquired for untreated, AS-PDT and TU-PDT absorbers, the spectrum is acquired in each case without any sputtering. (b) N-1s core level spectrum of TU-PDT absorber with three different etching times 0 second, 180 seconds and 1080 seconds to acquire information at different depths.

*Table S3.* Atomic ratio of [Cu] to [In] calculated using different energy lines of KCN etched CuInS$_2$ absorbers: untreated, AS-PDT and TU-PDT.

|  |  | [Cu]/[In] |
| --- | --- | --- |

| Sample Identifier | sputter time (s) | Cu 3p and In 3d | Cu 2p and In 3d | Cu 3p and In 4d |
|---|---|---|---|---|
| Untreated | 0 | 0.71 | 0.37 | 0.73 |
| Untreated | 180 | 0.83 | 0.65 | 0.90 |
| Untreated | 1080 | 0.94 | 0.77 | 1.02 |
| AS-PDT | 0 | 0.73 | 0.42 | 0.75 |
| AS-PDT | 180 | 0.84 | 0.66 | 0.91 |
| AS-PDT | 1080 | 0.93 | 0.76 | 1.01 |
| TU-PDT | 0 | 0.86 | 0.29 | 0.53 |
| TU-PDT | 180 | 0.80 | 0.54 | 0.76 |
| TU-PDT | 1080 | 0.93 | 0.72 | 0.97 |